\documentclass{nature}

\title{Titanic Magnetoresistance in WTe$_2$}

\author{Mazhar N. Ali$^1$, Jun Xiong$^2$, Steven Flynn$^1$, Quinn Gibson$^1$, Leslie Schoop$^1$, Neel Haldolaarachchige$^1$, N. P. Ong$^2$, Jing Tao$^3$ \& R. J. Cava$^1$}

\begin{document}

\maketitle

\begin{affiliations}
 \item Department of Chemistry, Princeton University, Princeton New Jersey 08544, USA.
 \item Joseph Henry Laboratories and Department of Physics, Princeton University, Princeton New Jersey 08544, USA.
 \item Department of Physics, Brookhaven National Laboratory, Upton, New York, 11973, USA.
\end{affiliations}

\begin{abstract}
\indent{}Magnetoresistance is the change of a material's electrical resistance in response to an applied magnetic field. In addition to its intrinsic scientific interest, it is a technologically important property, placing it in ``Pasteur's quadrant'' of research value: materials with large magnetorsistance have found use as magnetic sensors\cite{lenz1990review}, in magnetic memory\cite{Magmemory}, hard drives\cite{daughton1999gmr}, transistors\cite{PhysRevLett.97.077201}, and are the subject of frequent study in the field of spintronics\cite{wolf2001spintronics, SpintronicReview}. Here we report the observation of an extremely large one-dimensional positive magnetoresistance (XMR) in the layered transition metal dichalcogenide (TMD) WTe$_2$; 452,700\% at 4.5 Kelvin in a magnetic field of 14.7 Tesla, and 2.5 million\% at 0.4 Kelvin in 45 Tesla, with no saturation. The XMR is highly anisotropic, maximized in the crystallographic direction where small pockets of holes and electrons are found in the electronic structure. The determination of the origin of this effect and the fabrication of nanostructures and devices based on the XMR of WTe$_2$ will represent a significant new direction in the study and uses of magnetoresistivity.
\end{abstract}

\indent{} Large magnetoresistance (MR) is an uncommon property, mostly displayed by magnetic compounds. Giant Magnetoresistance (GMR)\cite{von1993giant, GMRpap} and Colossal Magnetoresistance (CMR)\cite{CMRcava, CMRpap}, are exhibited by thin film metals and manganese-based perovskites, for example. In contrast, ordinary magnetoresistance, a relatively weak effect, is commonly found in nonmagnetic compounds and elements\cite{Yang21051999}. Magnetic materials typically show negative magnetoresistances (where MR is defined as $\frac{\rho(H)-\rho(0)}{\rho(0)}$, is typically reported as a percent figure, and $\rho(H)$ is the resistivity in an applied magnetic field, H). Positive magnetoresistance is seen in both metals and semiconductors, where it is usually at the level of a few percent for metals; semiconducting silver chalcogenides show magnetoresistances up to 350\%, comparable with CMR materials\cite{xu1997large}. For single carrier type semiconductors, the MR behaves like (1+$\mu$H$^2$) where $\mu$ is the carrier mobility; high mobility semiconductors can therefore exhibit relatively large effects\cite{Solin01092000}. 

\indent{} WTe$_2$ is a TMD crystallizing in a distorted version of the common layered MoS$_2$ structure type\cite{AYQ:AYQA04998}. TMDs are known to display many interesting properties, such as catalysis of chemical reactions\cite{li2011mos2}, charge density waves (CDW)\cite{PhysRevB.16.801}, superconductivity\cite{PhysRevB.5.895}, have been exfoliated to fabricate interesting nanosctructures\cite{TMDnanostruct, radisavljevic2011single}, and now, we report, can display extremely large magnetoresistance. In the layered TMD compounds, metal layers are sandwiched between adjacent chalcogenide layers; this dichalogenide sandwich stacks along the \textit{c}-axis of the hexagonal structure, with van der Waals bonding between layers. Due to this anisotropic bonding, layered TMDs are typically electronically two-dimensional. In WTe$_2$, however, there is an additional structural distortion $-$ tungsten chains are formed within the dichalcogenide layers along the \textit{a}-axis of the orthorhombic unit cell, making the compound structurally one-dimensional (Figure 1a). WTe$_2$ is semi-metallic and has previously been investigated for thermoelectric applications in solid solutions with WSe$_2$ and MoTe$_2$\cite{thermoelectpap}. 

\indent{}Here we report the discovery of an extremely large positive magnetoresistance (XMR) in WTe$_2$ of up to 452,700\% at 4.5 Kelvin in an applied field of 14.7 Tesla (T) when the current direction is along the tungsten chains (\textit{a}-axis) and the magnetic field is applied perpendicular to the dichalcogenide layers, along the \textit{c}-axis. The magnetoresistance is still increasing at 45 T, the highest field in our measurements, where it has a value of 2.5 million percent with no saturation. It is especially surprising that this XMR is present in a non-magnetic, non-semiconducting system.  WTe$_2$ has a highly anisotropic electronic structure, with small pockets of holes and electrons in the directions where the XMR is maximized. The XMR is very anisotropic; largest along the chain direction, and dropping by more than 90\% when the magnetic field is applied in other directions, making this the largest one-dimensional magnetoresistance ever reported. The effect becomes significant at temperatures below $\approx$ 150 K; with the temperature of the ``turn on'' increasing with the magnitude of the applied magnetic field. Electron diffraction patterns taken at low temperatures indicate that the origin of the observed effect is not linked to the onset of a charge density wave or a Peierls-like distortion in the tungsten chains.  

\indent{}The temperature dependent resistivity under various applied magnetic fields ($\mu_0$H up to 14.7 T) is presented in Figure 2 on both logarithmic (main panel) and linear scales (lower inset). In zero-field, the room temperature resistivity is 0.6 m$\Omega$cm and falls to 1.9 $\mu$$\Omega$cm by 2 K, yielding a RRR of $\approx$ 370. When a field is applied, the resistivity of the sample essentially follows the zero-field curve until it is cooled close to the ``turn on'' temperature, T$^*$, (taken as the minimum in the resistivity) below which the resistivity begins to dramatically increase. The magnetoresistance effect at low temperatures is extremely large, reaching 452,700\% by 4.5 K in a 14.7 T field. The ``turn on'' temperature is shifted to higher temperature (at the rate of $\approx$ 4.4 K/T, upper inset) as larger fields are applied, implying a competition between dominating scattering mechanisms (phonon scattering and an as-yet unidentified mechanism responsible for the XMR). TEM electron diffraction studies (Figure 1c) at low temperature in a $\approx$ 2.8 T field (the field in the TEM at the sample position) show no evidence of a structural phase transition or the onset of a charge density wave to accompany the onset of the XMR effect. 

\indent{} Figure 3a shows the field-dependence of the XMR at various temperatures. The upper inset is a zoom of the main panel to clarify the XMR at higher temperatures while the lower inset shows the Shubnikov de Haas (SdH) quantum oscillations. The high quality of the crystals is shown not only in the high RRR, but also by the SdH oscillations in Figure 3b. The oscillations have been extracted after fitting a 2nd order polynomial to the 4.5 K parallel field measurement and subtracting that as background. They become visible by 6 T, begin increasing in amplitude, and then become dampened around 10.5 T before reemerging with much larger amplitude near 12.3 T. This behavior may be due to the interference between oscillations arising from the hole and electron pockets (see below) as their waves beat against each other. Figure 3b shows the 4.5 K MR field-dependence on the angle of the applied field to c-axis. When the field is aligned parallel to the c-axis the XMR effect is maximized with an $\frac{\rho(14.7)-\rho(0)}{\rho(0)}$ = 4527 (452,700\%). As the field is rotated to align parallel to either the \textit{a}-axis or the \textit{b}-axis, the MR effect is greatly diminished and dies like the cosine of the angle; this large anisotropy could be due to the very anisotropic Fermi surface of WTe$_2$ and scattering rates. Measurements up to 45 T at 0.4 K show an XMR of $\frac{\rho(45)-\rho(0)}{\rho(0)}$ $\approx$ 25,000 (MR $\approx$ 2,500,000\%) with strong quantum oscillations and still no resistivity saturation (Figure 3c). The magnetic field dependence of the resistivity is close to quadratic at intermediate fields (1-11 T), but displays a different field dependence at both higher and lower applied magnetic fields.

\indent{}Our electronic structure calculations show WTe$_2$ to be a semimetal where the valence band and conduction bands barely cross the Fermi energy at different places in the Brillouin zone (Figure 4a), an electronic structure that is reminiscent of that of the excitonic insulator TiSe$_2$\cite{PhysRevB.61.16213}. The detailed shape of the Fermi surface is (Figure 4b) very sensitive to the position of the Fermi level. The band structure shows the electron and hole pockets along the $\Gamma$ - X direction that make WTe$_2$ a semimetal, as well as a potential second set of electron and hole pockets forming along Z - U. The $\Gamma$ - X direction in reciprocal space corresponds to the \textit{a}-axis in real space, or along the tungsten chains. Since the Z - U direction is parallel to the $\Gamma$ - X direction, but shifted along k$_z$ into the perpendicular face of the Brillouin zone, this potential second crossing would change the pockets into tubes in the Fermi surface. Future ARPES study and more detailed transport analysis and characterization of the quantum oscillations will be needed to determine the details of the Fermi surface and form a basis for understanding the observed XMR. 

\indent{}With a positive magnetoresistance this large, and the one-dimensional behavior of the XMR there are few comparable systems to WTe$_2$. The closest case appears to be high purity Bismuth, a semimetal with pockets in its Fermi surface,\cite{mangez1976transport, giura1967magnetoacoustic, Yang21051999} which also shows extremely large positive magnetoresistance\cite{PhysRev.91.1060} comparable in magnitude to what is reported here. WTe$_2$ is different from Bi in a number of fundamental ways; for example, WTe$_2$ is structurally one-dimensional and shows an extreme one-dimensional anisotropy in its XMR, the XMR in WTe$_2$ has a sharper ``turn on'' temperature than is seen in Bi, WTe$_2$ is a semimetal with extremely small overlap between valence band and conduction band states, resembling an excitonic insulator\cite{PhysRevB.61.16213}, and ordinary purity WTe$_2$ shows the effect. Particularly important from a materials development perspective is the fact that WTe$_2$ is a layered TMD that can be easily exfoliated and therefore form the basis for thin films and advanced nanostructures similar to MoS$_2$\cite{radisavljevic2011single}. 

\indent{}The single crystals made in this study were crudely exfoliated by using double sided tape, and thicknesses down to a few microns were easily achieved. Evaporation growth and subsequent annealing to make single crystal thin films may enhance the MR effect seen here due to higher crystal quality. Hybrid structures of various kinds, such as layering WTe$_2$ with magnetic films, combined with the XMR effect may be useful in devices such as highly sensitive low temperature magnetic field sensors or high field temperature sensors in cryogenics. In particular, the one-dimensional aspect of the anisotropic MR in WTe$_2$ may be useful in low temperature magnetic field sensing and, especially, orienting. In fact, recently it was reported that below 20 K or above 5 T\cite{ekin2006experimental, Sensors2011} the materials currently used for temperature or field measurements are prone to large degrees of error. In contrast, however, this regime is actually where WTe$_2$ performs best. The ease with which this system can be exfoliated as well as the effect that even small changes in the Fermi level may have on the properties makes it an ideal candidate for electron gating experiments. Careful chemical doping and intercalation of WTe$_2$ may also be key in elucidating the cause of the XMR and potentially unlocking further properties of interest.


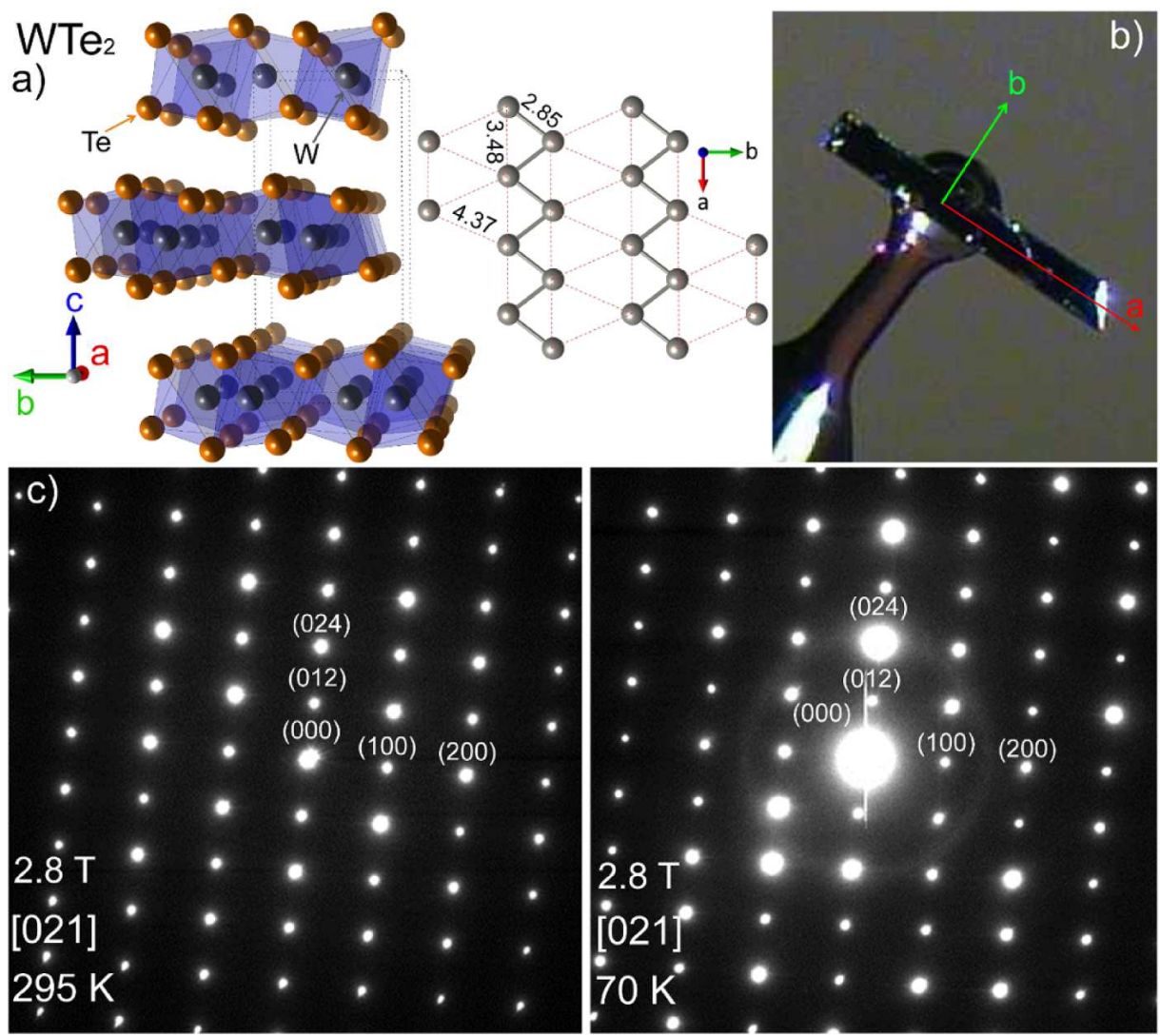
\begin{figure}
	\caption{\scriptsize{\textbf{(color online):} Structural considerations. Panel a) the crystal structure of WTe$_2$, showing the layered nature, typical of what is seen for transition metal dichalcogenides, and also the chains of W atoms along the \textit{a}-axis distorting the ideal hexagonal net. All distances are in \AA. Panel b) a typical crystal of WTe$_2$, crystallographic directions marked. The XMR exists when the current (I) flows along \textit{a} and the field is parallel to \textit{c} (see below). Panel c) electron diffraction images looking down the [021] zone showing the reciprocal lattice at room temperature and low temperature. The data indicate that there has been no structural transition in WTe$_2$ upon cooling (effective magnetic field at the sample in the TEM is about 2.8 T).}}
	\label{Figure_1}
\end{figure}


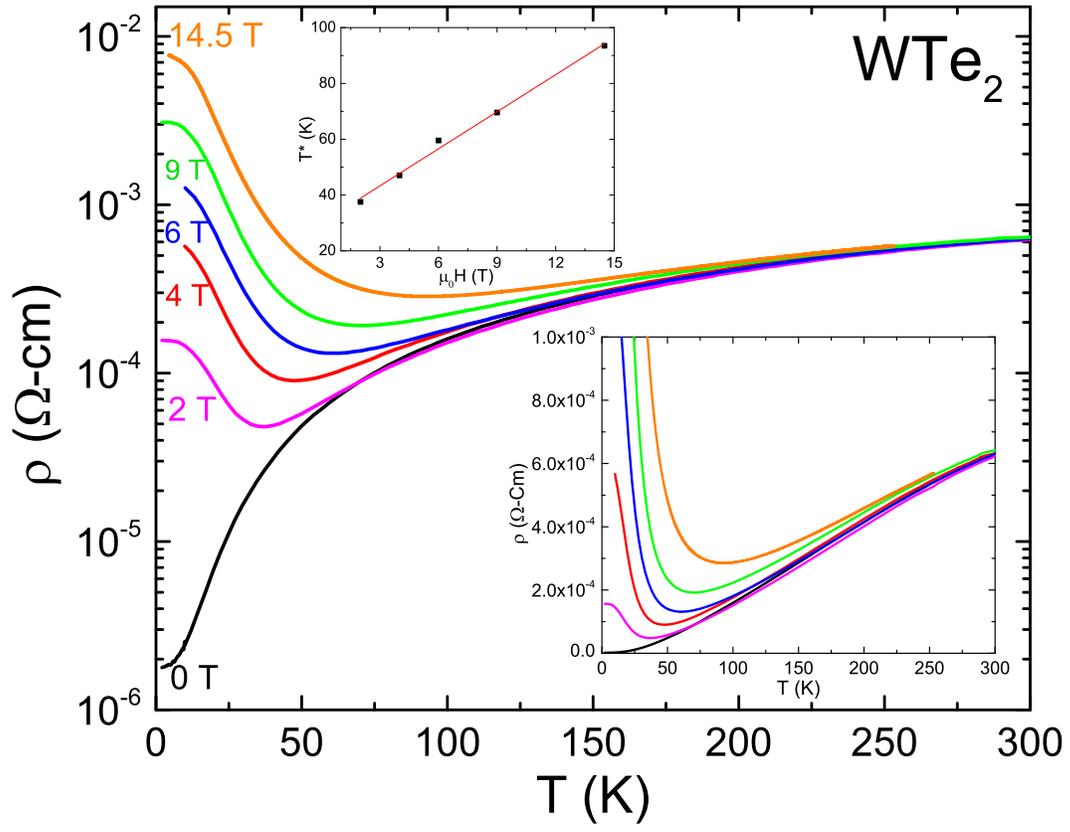
\begin{figure}
	\caption{\scriptsize{\textbf{(color online):} The temperature and field dependence of the extremely large magnetoresistance in WTe$_2$. Main panel - log of resistivity vs temperature. Lower inset, resistivity vs. temperature showing turn on of the effect. T* is defined as the temperature where the resistivity is a minimum - an approximation of the turn on temperature of the XMR. Upper inset, the linear dependence of T* on magnetic field, the slope is equal to 4.4 K/T.}}
	\label{Figure_2}
\end{figure}


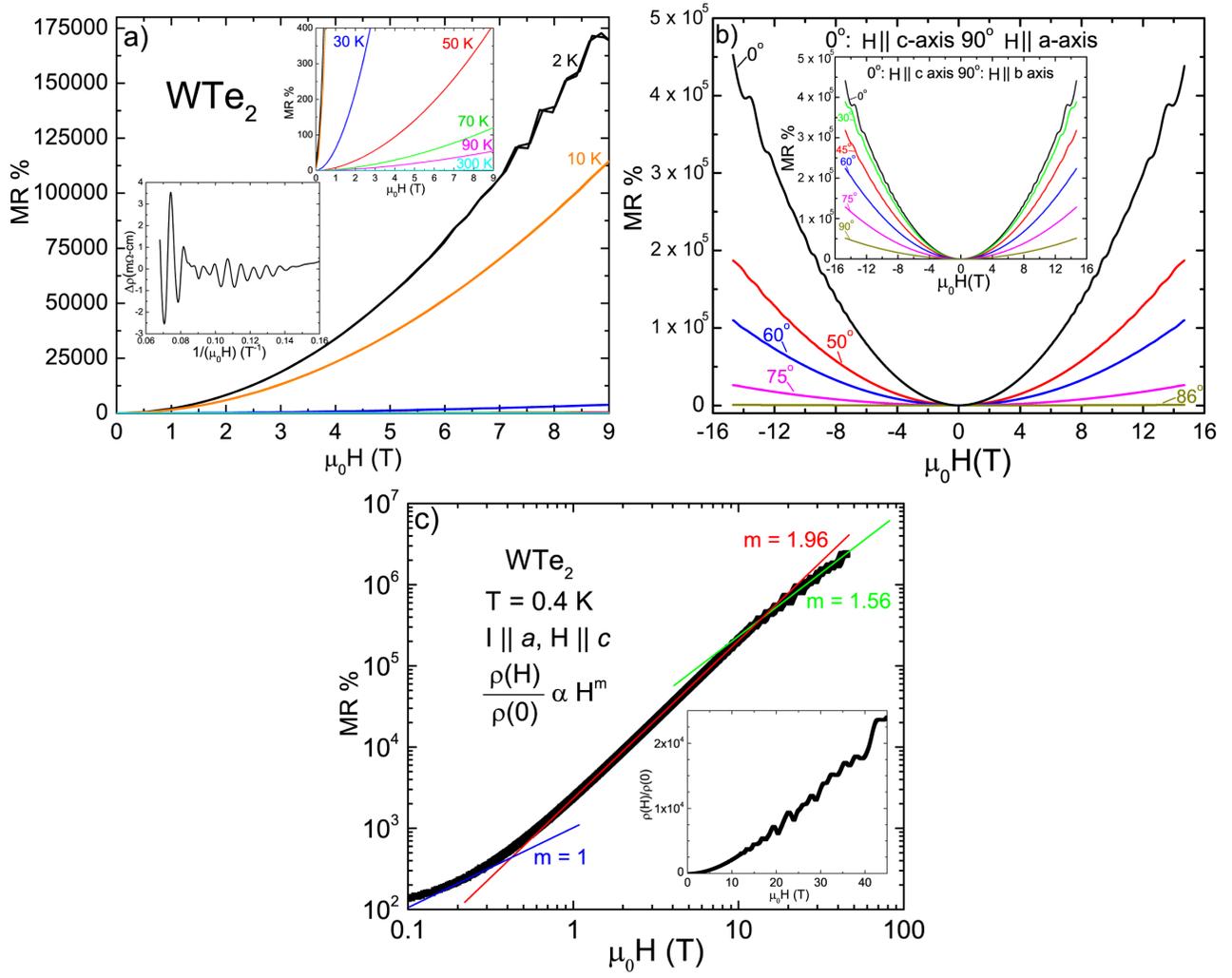
\begin{figure}
	\caption{\scriptsize{\textbf{(color online):} Field and angular dependence of the XMR in WTe$_2$. Panel a) the field dependence of the XMR in WTe$_2$ with the current along the W-W chains and the applied field parallel to the \textit{c}-axis from 0 - 9 T at representative temperatures. Upper inset shows detail of the magnetoresistance at higher temperature. Lower inset shows the detail of the quantum oscillations at 4.5 K. This demonstrates the high quality of the crystal and suggests that carriers in the hole and electron pockets (Figure 4) `beat' against each other at intermediate field values. Panel b) the angular dependence of the XMR at 4.5 K shows that the effect is one-dimensional; it is maximized when H is parallel to \textit{c} and goes to 0 when H is perpendicular to \textit{c}. The main panel shows the MR as the applied field is rotated to be parallel to \textit{a} and the inset shows the same effect when the field is rotated to be parallel to \textit{b}. Panel c) The XMR of WTe$_2$ up to 45 T. The m = 1.96 line is from a fit of $\frac{\rho(H)}{\rho(0)}$ $\alpha$ H$^m$ from 1 - 11 T, m = 1.56 line is from a fit of $\frac{\rho(H)}{\rho(0)}$ $\alpha$ H$^m$ from 11 - 45 T. m = 1 at low field is a guide to the eye.}}
	\label{Figure_3}
\end{figure}


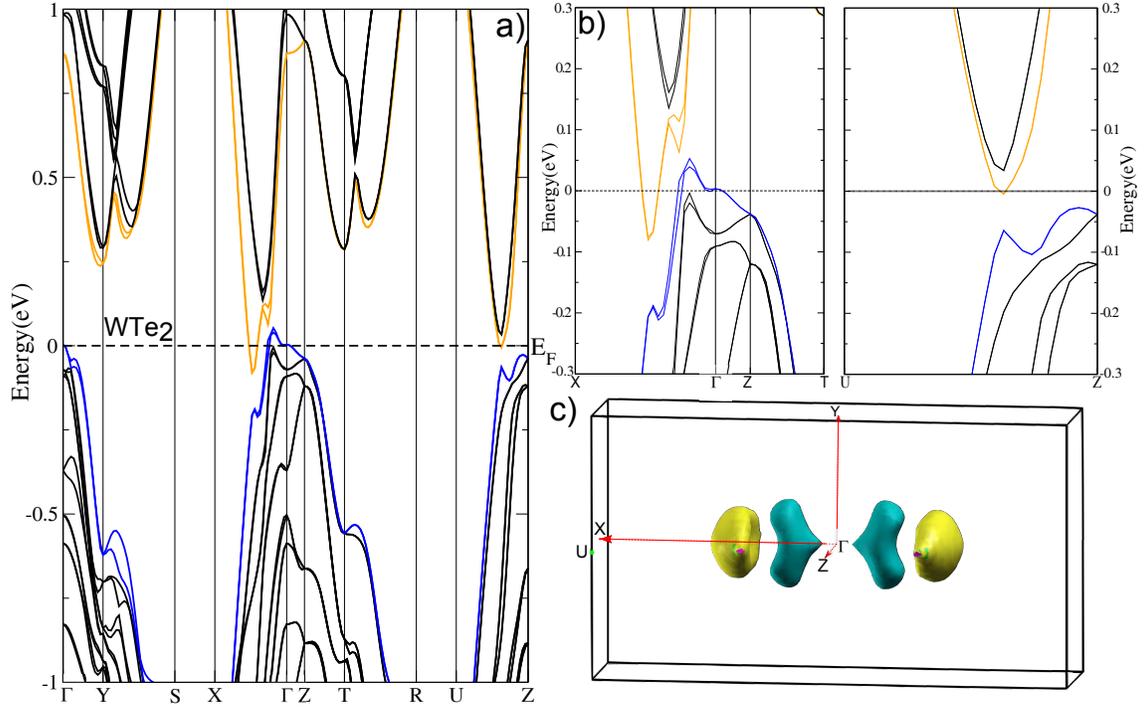
\begin{figure}
	\caption{\scriptsize{\textbf{(color online):} The electronic structure of WTe$_2$, calculated including spin orbit coupling. a) The energy vs. wavevector relationships for high symmetry directions in the orthorhombic brillouin zone. Note the semimetal character due to the crossing of the hole and electron bands near $\Gamma$. b) Detail of the calculated electronic structure in the $\Gamma$ - X and $\Gamma$ - Z directions and similar detail showing the possible crossing of the states between Z and U. c) The Fermi surface of WTe$_2$ showing electron (yellow) and hole (blue) pockets displaced from the $\Gamma$ point and aligned along the chain direction.}}
	\label{Figure_4}
\end{figure}

\begin{methods}
\indent{}High quality single crystals of WTe$_2$ were grown via Br$_2$ vapor transport. Tungsten powder was ground together with purified Tellurium, pressed into a pellet, and heated in an evacuated quartz tube at 700 $^{\circ}{\rm C}$, homogenized, then reheated at 750 $^{\circ}{\rm C}$ for 2 days each. This final pellet was ground into a fine powder and a temperature gradient of 100 $^{\circ}{\rm C}$ between 750 $^{\circ}{\rm C}$ - 650 $^{\circ}{\rm C}$ was used for crystal growth, with a Br$_2$ concentration of $\approx$ 3 mg/ml in a sealed quartz tube for 1 week. Optimal crystals were obtained under these conditions; crystals grown at higher temperatures showed substantially lower residual resistivity ratios (RRR) $\frac{\rho(300 K)}{\rho(2 K)}$ and degraded magnetoresistance. The need to employ low temperature synthesis to avoid defect formation that degrades properties is frequently observed in TMDs, for example in TiSe$_2$\cite{PhysRevB.14.4321}. The crystals grew as thin ribbons (Figure 1b), with the long direction being the W-W chain direction and the larger flat faces being perpendicular to the stacking direction of the layers.

\indent{}WTe$_2$ crystals were structurally and chemically characterized by powder-XRD to confirm bulk purity, single crystal XRD to determine crystal growth orientation, SEM-EDX for chemical analysis, and TEM to search for a low temperature phase transition. For general electronic characterization, SQUID magnetometer measurements revealed weak diamagnetism typical of a metal, and thermopower measurements confirm a previously reported n-p type crossover at 65 K in a 0 T field and n-type Hall effect behavior down to 2 K \cite{WTe2_elect}. In resistivity measurements, electrical anisotropy was found; the tungsten chain direction (the \textit{a}-axis) had the lowest resistivity.

\indent{}Powder x-ray diffraction patterns were collected using Cu K$\alpha$ radiation on a Bruker D8 Focus diffractometer with a graphite monochromator. Single crystal x-ray diffraction data was collected on a Bruker APEX II using Mo K$\alpha$ radiation ($\lambda$ = 0.71073 \AA) at 100 K. Scanning electron microscopy and energy dispersive x-ray analysis were carried out on a FEI Quanta 200 FEG Environmental-SEM and was used to determine the composition of the crystals. Electron diffraction was carried out in a JEOL 3000F transmission electron microscope equipped with a Gatan liquid-helium cooling stage.

\indent{}The magnetoresistance of WTe$_2$ samples was measured using the 4-point probe method in a Quantum Design PPMS and with a Delta-mode method by a Keithley 6221 current source meter and a 2182A nanovoltmeter. The high-field dependent data were taken at 4.5 K up to 14.7 T in an American Magnetics superconducting magnet. Resistivity measurements up to 45 T were performed at the National Magnet Laboratory. Magnetic susceptibility was measured as a function of temperature in the 2 - 300 K range at applied fields of 0.3 and 3 Tesla. 

\indent{}The electronic structure calculations were performed in the framework of density functional theory using the \textsc{wien2k}\cite{blaha2001} code with a full-potential linearized augmented plane-wave and local orbitals [FP-LAPW + lo] basis\cite{singh2006} together with the Perdew Burke Ernzerhof (PBE) parametrization of the generalized gradient approximation (GGA) as the exchange-correlation functional. The plane wave cut-off parameter R$_{MT}$K$_{MAX}$ was set to 8 and the Brillouin zone was sampled by 10000 k-points. The convergence was confirmed by increasing both R$_{MT}$K$_{MAX}$ and the number of k points until no change in the total energy was observed. The Fermi surface was plotted with the program \textit{Xcrysden}.

\end{methods}

\bibliography{Lit}

\begin{addendum}
 \item This research was supported by the Army Research Office, grant W911NF-12-1-0461, and the NSF MRSEC Program Grant DMR-0819860. The authors would like to thank Emanuel Tutuc for inquiring about WTe$_2$. The National Magnet Laboratory is supported by the National Science Foundation Cooperative Agreement No. DMR-1157490, the State of Florida, and the U.S. Department of Energy. The electron microscopy study at Brookhaven National Laboratory was supported by the US Department of Energy, Basic Energy Sciences, under contract DE-AC02-98CH10886.
 \item[Author Contributions] Mazhar Ali and Steven Flynn grew the samples and measured the resistivity as a function of temperature and field. Neel Haldolaarachchige assisted with those measurements. Jun Xiong also measured the resistivity as at high fields and various angles. Quinn Gibson and Leslie Schoop calculated the electronic structure. Jing Tao measured electron diffraction. N. P. Ong and R. J. Cava are the principal investigators.
 \item[Competing Interests] The authors declare that they have no
competing financial interests.
 \item[Correspondence] Correspondence and requests for materials
should be addressed to Mazhar N. Ali~(email: mnali@princeton.edu).
\end{addendum}

\end{document}